\documentclass[runningheads]{ceurart}
\usepackage[T1]{fontenc}
\usepackage{graphicx}
\usepackage{amsmath, amssymb}
\usepackage{algorithm}
\usepackage[noend]{algpseudocode}
\usepackage{xcolor}
\begin{document}
\title{From Data Footprints to Researcher Fingerprints: Temporal Ego-centric Research Group Detection in Scientific Knowledge Graphs}
%
%
\author[1]{Simone Angioni}[%
orcid=0000-0002-5193-7851,
email=simone.angioni@isti.cnr.it,
]
\fnmark[1]
\address[1]{Institute of Information Science and Technologies ``Alessandro Faedo'' (ISTI), National Research Council of Italy (CNR), Pisa, Italy}
\address[2]{OpenAIRE AMKE, Athens, Greece}

\author[1]{Miriam Baglioni}[%
orcid=0000-0002-2273-9004,
email=miriam.baglioni@isti.cnr.it
]
\fnmark[1]

\author[1]{Andrea Mannocci}[%
orcid=0000-0002-5193-7851,
email=andrea.mannocci@isti.cnr.it
]
\cormark[1]
\fnmark[1]

\author[1,2]{Paolo Manghi}[%
orcid=0000-0001-7291-3210,
email=paolo.manghi@isti.cnr.it
]

\cortext[1]{Corresponding author.}
\fntext[1]{These authors contributed equally.}
%
%
%
\copyrightyear{2026}
\copyrightclause{Copyright for this paper by its authors. Use permitted under Creative Commons License Attribution 4.0 International (CC BY 4.0).}
\conference{6th International Workshop on Scientific Knowledge: Representation, Discovery, and Assessment, Oct 2026, Bari, Italy}

\begin{abstract}
Understanding who researchers collaborate with, when and how their research groups form, and how these evolve over time is a foundational prerequisite for constructing multidimensional, evidence-based researcher fingerprints -- a requirement that has become increasingly urgent in the context of ongoing reforms in research evaluation advocated by initiatives such as DORA, the Leiden Manifesto, and CoARA. 
In this paper, we propose an egocentric approach for fine-grained research group detection, operating at the scale of individual researchers' ego networks, as extracted from Scientific Knowledge Graphs (SKGs), and requiring the ego node to be present in all identified research groups -- a constraint that is not merely technical but reflects a commitment to researcher-centred analysis. 
Our approach builds on the Leiden algorithm as its computational core and is grounded in SKGs, which provide the heterogeneous, open scholarly metadata necessary to trace collaboration patterns at scale in a reproducible and transparent manner. 
We evaluate the proposed method through a user evaluation involving 10 researchers with diverse backgrounds and career stages, demonstrating its ability to surface meaningful and interpretable collaboration structures. 
The research groups identified constitute the structural scaffold upon which richer characterisations of scholarly identity -- including roles, seniority, contribution types, and publication patterns -- can be built in subsequent work, contributing to the broader goal of fairer and more transparent research assessment.
\end{abstract}

\maketitle

\section{Introduction}
The landscape of research evaluation is undergoing a paradigm shift. 
Driven by global initiatives such as DORA~\cite{dora2013}, the Leiden Manifesto~\cite{dianahicks2015}, and, more recently, CoARA~\cite{arentoft2022}, which advocates a transition from purely quantitative, publication-centric assessment to a broader, more qualitative evaluation of researchers' contributions, new approaches to recognising scholarly merit are emerging~\cite{johansen2021,scope2023,oneill2026}. 
Researchers are no longer to be judged solely by problematic metrics and indicators, such as the h-index or the prestige of the journals and conferences in which they publish, but by the broader impact of their work on advancing all aspects of science.
All research contributions should be assessed on the basis of their intrinsic quality and their potential to be reused, built upon, and integrated by others -- fostering a more collaborative and cumulative mode of scientific progress.

Yet qualitative peer review, while remaining the gold standard in the eyes of the research evaluation reform, is resource-intensive and difficult to sustain at scale. 
Therefore, in a robust and fair evaluation framework, peer review must be complemented by novel, comprehensive, data-driven insights that capture the full scope of a researcher's activity.

Much like a footprint, each researcher leaves a distinctive and traceable pattern of contributions across the scholarly record -- datasets deposited, software released, collaborations established, citations given and received across disciplines, and knowledge transferred beyond academia. 
Taken together, these signals compose a research fingerprint: a multidimensional, evidence-based portrait of a scholar's engagement with Open Science practices and the cumulative reach of their work.

Constructing such a fingerprint and understanding its evolution, however, cannot proceed in a single step. 
The signals that compose it, such as contribution types, seniority and leadership~\cite{dino2020,zhou2025a}, scientific production patterns~\cite{dino2020,yao2025,jaramillo2023}, impact, citation reach, academic mobility, and knowledge transfer across disciplinary boundaries and towards the industry~\cite{salatino2026}, acquire their full interpretive power only when situated within the collaborative contexts in which a researcher operates throughout time.
Understanding who researchers work with, when and how their research groups form, and how these groups develop, is therefore not merely one dimension of the fingerprint among others, but rather the foundational layer upon which all remaining dimensions rest.
This paper addresses precisely this phase by proposing a method to identify and trace research groups at the level of the individual researcher, establishing the structural scaffold without which finer-grained characterisations of role, publication and impact dynamics, and scholarly identity cannot be meaningfully constructed.

The empirical backbone of this analysis is provided by Scientific Knowledge Graphs~\cite{auer2018} (SKGs): large-scale, open infrastructures that integrate heterogeneous scholarly metadata -- authors, publications, datasets, software, funding, and their semantic relations -- into a unified representation of the research landscape. 
SKGs such as the OpenAIRE Graph~\cite{manghi_2025_17725827} offer the coverage, richness, and openness necessary to trace collaboration patterns at scale and to ground individual research fingerprints in reproducible, transparent evidence -- an approach fully consistent with the Open Science principles~\cite{posiadopters2025} that motivate the evaluation reform agenda described above.

From a methodological standpoint, detecting research groups within such graphs poses specific challenges. 
State-of-the-art community detection approaches typically operate on large-scale co-authorship networks, partitioning them into clusters based on macro-level topological patterns and optimising for global properties such as modularity -- without regard for the perspective of the individual researcher. 
Notably, algorithms such as DEMON~\cite{coscia2012} and ANGEL~\cite{rossetti2020} have been proposed for general social networks, but their application to scientific collaboration -- and the specific constraint of ego membership in all detected subgroups -- remains unexplored.

The approach proposed here departs from this paradigm by operating at the scale of the ego network: rather than partitioning the global graph, the algorithm analyses the local neighbourhood of a researcher and identifies collaboration subgroups, ensuring the ego node is present in each.
This constraint is not merely technical, as it reflects a conceptual commitment to keeping the individual researcher at the centre of the analysis, ensuring that the groups identified are meaningful with respect to that researcher's specific collaborative experience and evolving role.

Against this backdrop, the contribution of this paper is twofold. 
First, it introduces an egocentric approach for fine-grained research group detection -- operating at the scale of individual researchers' ego networks and enforcing ego membership in all identified subgroups -- building on the Leiden algorithm for community detection~\cite{traag2019louvain} as its computational core.
Second, it validates the proposed approach with a user evaluation involving 10 researchers with diverse backgrounds and career stages.
Together, these contributions establish the structural scaffold upon which subsequent work will build richer, multidimensional researcher fingerprints in service of more transparent and responsible research assessment.

The remainder of this paper is structured as follows. Section~\ref{sec:soa} reviews the state of the art in both community detection and research group identification. Section~\ref{sec:data} describes the dataset at the basis of this analysis, while Section~\ref{sec:methods} introduces the methodology for the construction of the ego network and the algorithm for iterative research group identification. Section~\ref{sec:tool} provides an overview of the tool, while Section~\ref{sec:eval} presents a user evaluation of the proposed approach. Finally, Section~\ref{sec:conclusion} addresses the current limitations and outlines future research directions.

\section{Related Works}\label{sec:soa}
Community detection is a central task in network science and aims to identify groups of nodes that are more densely connected internally than externally. 
Over the last two decades, it has become one of the main analytical tools for studying scientific collaboration and co-authorship networks, in which communities are typically interpreted as cohesive collaboration structures, topical clusters, or partially institutionalised scientific circles. 
Recent reviews confirm that the field now includes a broad spectrum of approaches, ranging from modularity optimisation and spectral methods to probabilistic models, overlapping-community methods, and graph representation learning~\cite{khawaja2024,li_review2024,diboune_survey2024,fortunato2022}.

Among graph-partitioning methods, modularity-based algorithms remain especially influential because they combine interpretability with scalability on large networks. 
Louvain was long the standard~\cite{blondel2008}, but Leiden~\cite{traag2019louvain} is now widely preferred because it improves partition refinement and guarantees better connected communities. 
This is particularly important in co-authorship networks, where poorly connected or fragmented partitions can distort the interpretation of collaboration structures.

At the same time, the literature has made clear that there is no universally best definition of community. 
As emphasised in recent surveys, community detection now includes several distinct problem settings: disjoint vs overlapping communities~\cite{coscia2012,rossetti2020}, static vs dynamic communities~\cite{cazabet2017,zhuang2019,seifikar2020}, and topology-only vs attributed or representation-learning-based methods. All these aspects are relevant to scholarly collaboration, as researchers often participate in multiple, partially overlapping, and evolving groups. 
As such, a single hard partition of a graph may therefore be analytically useful but still insufficient to capture the full-fledged diversity of an author’s collaborative environment~\cite{fortunato2022,li_review2024,singh2025}.
Other works used dynamic knowledge graphs to detect emerging communities in scientific data~\cite{aparicio2024}, and network embeddings to detect communities~\cite{zhou2025}.

Dynamic and temporal community detection has become a major research direction in the field~\cite{rossetti2019,seifikar2020}.
Instead of analysing a collaboration network as a static graph, newer methods model the evolution of communities across time slices or temporal graphs \cite{rossetti2020,bazzi2016}. 
Recent work has explored dynamic modularity, contrastive learning, graph convolutional approaches, and temporal uncertainty-aware methods to better capture how communities emerge, split, merge, and disappear~\cite{mucha2010,bazzi2016,li2023,khawaja2024}. This direction is especially relevant for scholarly data because scientific careers are intrinsically longitudinal, and research groups are rarely fixed over time.

Since research group detection is a more specific task than general community detection, the literature is comparatively less consolidated and often overlaps with studies on scientific teams, co-authorship communities, and collaboration structures.
Perianes-Rodríguez et al.~\cite{perianes-rodriguez2010} represent one of the earliest works for research group detection from co-authorship data. Their work proposes a method for detecting, identifying, and visualising research groups by constructing co-authorship networks and applying structural analysis together with factorial analysis of the raw co-authorship matrix. Research groups are identified according to similarities in researchers' choices of co-authors, and the method is applied to the departments of Carlos III University of Madrid.
Reyes-Gonzalez et al.~\cite{reyes-gonzalez2016} proposed a method that combines co-authorship and citation analysis to identify research groups as self-organising units of scientific production. Their idea was that researchers organise around key collaborators and that research groups should be treated as self-generated structures rather than as predefined administrative units. In addition to co-authorship relations, they use citation-based information to characterise and compare the productivity and performance of research groups. The method is evaluated on research groups in Physics, Applied Physics, Condensed Matter, Materials Science, and Optics in leading Mexican institutions. 
Cheng et al.~\cite{cheng2024} recently proposed a method for identifying different types of university research teams. Their approach distinguishes among project-based research teams, individual-based research teams, backbone-based research groups, and representative research groups, using information about projects, contributions, collaboration, and similarity. They demonstrated that research-team detection is evolving beyond the identification of a single generic collaboration cluster, and that different definitions of research teams may require different identification criteria.
However, while highly relevant to our topic, these works mainly focus on detecting research groups within institutional or disciplinary settings, classifying different types of university research teams, or assessing the performance of self-organising groups around key researchers. In contrast, our work aims to reconstruct all the temporally evolving research groups associated with a focal author, independently of whether they correspond to a single institution, discipline, or formal team structure.

Currently, existing approaches typically identify communities or research groups based on a single partition of the collaboration graph, a local cluster surrounding a focal scholar, or a team constructed for a specific period. 
As a result, these methods are effective at detecting cohesive structures within a network but are less capable of reconstructing the full range of research groups a single researcher has engaged with throughout their career. 
This limitation is particularly pressing for interdisciplinary, mobile, or senior researchers, whose collaboration history often spans multiple, partially overlapping groups, institutions, and time periods.

Our work differs from the current state of the art in three ways. 
First, it is explicitly ego-centred; the objective is not to globally partition the entire graph, but to reconstruct the lineage of a single researcher's collaboration environment. 
Second, communities are not treated as the final output; instead, they serve as an intermediate structural organisation that helps contextualise relevant collaborators. 
Third, we determine research group membership based on yearly relevance to the author’s production and track this information across years. 
This approach enables us to identify multiple, temporally persistent research groups rather than relying on a single static cluster or a one-time partitioning strategy.


\section{Data}\label{sec:data}
The empirical foundation of this work is the OpenAIRE Graph (hereafter the Graph), one of the largest open scientific knowledge graphs available today. 
It provides a comprehensive view of the research lifecycle, treating publications, research data, research software, and other research products as first-class citizens, and connecting them through semantic relationships, including citations, affiliations, project participation, and funding links.
The Graph aggregates, deduplicates, and enriches metadata from over 2,100 direct sources, rising to approximately 155,000 when indirect sources are included.
Many different types of sources are considered, ranging from institutional and thematic repositories to well-established services such as CrossRef, Datacite, and PubMed.
Updated once per month and released every six months as an open snapshot on Zenodo under a CC-BY licence~\cite{manghi_2025_17725827} and as a dataset on Google Big Query\footnote{OpenAIRE Cloud Access, \url{https://graph.openaire.eu/docs/cloud-access}}, the Graph aggregates hundreds of millions of metadata records from sources trusted by the scientific community. 

The choice of the OpenAIRE Graph over alternative SKGs is motivated by its open licence, its treatment of non-traditional outputs as first-class research products, and its transparent, community-governed curation process -- all essential properties for reproducible and auditable researcher fingerprint construction.
For the purposes of this work, given a researcher selected as the ego node, we used their ORCID identifier to extract all associated research products from the OpenAIRE Graph $10.6.0$ snapshot (end of 2025), including publications, datasets, software, and other research product types. 
It is worth noting that, for a given ORCID identifier, the OpenAIRE Graph comprehends those explicitly verified on the researcher's ORCID profile (i.e., listed in the ``Works'' section), as well as those inferred by OpenAIRE through dedicated enrichment algorithms\footnote{\url{https://graph.openaire.eu/docs/category/deduction--propagation}}.

\section{Methodology}\label{sec:methods}
In the following section, we mathematically ground the construction of the ego network, starting from an ego node and its research products retrieved from the OpenAIRE Graph by its ORCID identifier (Section~\ref{sec:ego}).
Then, we describe the iterative algorithm for detecting the communities the ego node belongs to (Section~\ref{sec:community-detection}), and ultimately, we proceed with describing the logic behind the final step of the proposed method in charge of the construction of the ego's research groups (Section~\ref{sec:research-group-detection}).

\subsection{Ego Network Construction}\label{sec:ego}
To analyse how the communities a focal author participates in are actually organised, a standard ego network is often insufficient. In the classical construction~\cite{Perry_Pescosolido_Borgatti_2018}, the ego network of an author $a_{ego}$ includes the ego node and its directly connected collaborators, but the edge weights only retain collaborations in which the ego itself participates. 
As a consequence, the resulting graph provides only a partial view of the relationships among the members of the ego node’s neighbourhood~\cite{crossley2015social}.
This limitation is particularly relevant when the goal is not merely to identify an author's collaborators, but to understand how those collaborators are structured into communities. 
Two co-authors may both belong to the neighbourhood of $a_{ego}$, yet their mutual relation may be significantly underestimated -- or even absent in weighted terms -- if their collaboration mostly occurs independently of the ego. 
In such cases, the standard ego network does not preserve the actual connectivity pattern among neighbouring authors and may therefore distort the internal composition of communities.

For this reason, we used the induced weighted ego network to preserve the same set of authors in the ego neighbourhood while inheriting the true relations among them from the global graph. 
In other words, the induced weighted ego network enriches the ego-centred network by restoring the original pairwise connections among all the authors it contains. 
This allows the local network to reflect how the members of the ego’s neighbourhood are genuinely connected, regardless of whether those collaborations involve the ego node directly.

Formally, let $P$ be a set of publications and $A$ a set of authors. We define $auth: P \to 2^A$ as the function that maps each publication to its set of authors.

We model the co-authorship network as a weighted graph:
\begin{equation*}
    G = (A, R, w)
\end{equation*}
Where:

$A$ is the set of authors; 
$R \subseteq A \times A $ is the set of co-authorship relations defined as
    \begin{equation*}
         (a_i, a_j) \in R  \Leftrightarrow 
        \exists\, p \in P : a_i \in auth(p) 
        \wedge a_j \in auth(p) 
        \wedge a_i \neq a_j
    \end{equation*}
and $w: A \times A \to \mathbb{N} $ is the weight function, where $w(a_i,a_j)$ counts the number of publications co-authored by $a_i$ and $a_j$, defined as
    \begin{equation*}
        w(a_i, a_j) = 
        \bigl|\{p \in P \mid a_i \in auth(p) 
        \wedge a_j \in auth(p) \wedge a_i \neq a_j\}\bigr|
    \end{equation*}
Given an ego node $a_{ego} \in A$, we define the ego network $G_{ego}$ as the subgraph: 
\begin{equation*}
G_{ego} = (A_{ego}, R_{ego}, w_{ego})
\end{equation*}
where 
\begin{equation*}
A_{ego} = \{a_{ego}\} \cup \{a_i \in A \mid  (a_{ego},a_i) \in R\}  
\end{equation*}
is the ego network node set composed of the ego node and all its direct neighbours,
\begin{equation*}
        R_{ego} = \{(a_i, a_j) \in R \mid a_i \in A_{ego} \wedge a_j \in A_{ego} \}
\end{equation*}
is the relation set, and
\begin{equation*}
    w_{ego}(a_i, a_j) = \bigl|\{p \in P \mid a_i \in auth(p) \wedge a_j \in auth(p) \wedge a_{ego} \in auth(p) \wedge a_i \neq a_j \}\bigr|
\end{equation*} 
is the weight function.

The induced weighted ego network $G^{induced}_{ego}$ is the sub-graph of $G$ induced by $A_{ego}$, where edge weights count all co-authorship among nodes independently of the presence of $a_{ego}$
\begin{equation*}
    G^{induced}_{ego}=(A_{ego},R_{ego},w_{aug})
\end{equation*}
where 
\begin{equation*}
    w_{aug}(a_i, a_j) = \bigl|\{p\in P \mid a_i \in auth(p) \wedge a_j \in auth(p) 
    \wedge a_i \neq a_j
    \}\bigr| 
\end{equation*}


It is important to note that $G_{ego}$ and $G^{induced}_{ego}$  share the same node and edge sets, but differ in how edge weights are computed: in $G_{ego}$ only publications involving $a_{ego}$ contribute to the weight, whereas in  $G^{induced}_{ego}$ all co-authorships among nodes of $A_{ego}$ are counted.

\subsection{Community Detection}\label{sec:community-detection}
This step of the proposed method uses at its core the state-of-the-art Leiden algorithm for community detection and embeds it in a customised procedure addressing the particular requirements of the current setting.

Generally speaking, the Leiden algorithm optimises a partition of a graph by assigning each node to exactly one community according to a global quality function. When applied to an ego network, however, it operates only on the local neighbourhood of the ego node rather than on the full co-authorship graph. This has two important consequences. First, the relational context of each co-author is only partially observed, since all connections outside the ego network are excluded. Second, because Leiden produces a hard partition, every node is forced into exactly one community -- even when actual collaboration patterns are overlapping or ambiguous. As a result, weakly connected or multidisciplinary co-authors may be assigned to communities that would differ if the full global network were available. Communities identified in this setting should therefore be interpreted as \textit{local ego-centred communities} rather than globally optimal partitions.

A further limitation of a single application of Leiden is that it returns one partition of the ego network, which may conflate distinct research groups into a single community or suppress smaller but meaningful collaboration structures associated with the ego node. This is particularly problematic for senior or interdisciplinary researchers, whose collaboration history often spans multiple partially overlapping groups.

To address both limitations, we adopt a recursive application of Leiden, formalised in Algorithm~\ref{alg:community_extraction}. At each iteration, the community containing $a_{ego}$ is extracted and its members -- excluding the ego node itself -- are removed from the graph before the algorithm is run again. The first iteration captures the most prominent ego-centred community in the local structure; subsequent iterations progressively reveal additional research groups associated with $a_{ego}$ that would otherwise remain hidden within a single hard partition. This recursive peeling strategy allows the algorithm to reconstruct multiple, distinct collaboration groups across the ego's career, rather than collapsing them into a single static cluster.

\begin{algorithm}[t!]
    \caption{Community Extraction Algorithm}\label{alg:community_extraction}
    \begin{algorithmic}[1]
        \Procedure{communityExtraction}{}
        \State$G\gets G^{induced}_{ego} = (A, R, w)$
        \State$communities \gets \emptyset $
        \Repeat
            \State $C\gets leiden(G)$ \Comment{Compute Leiden Algorithm on the current Graph}
            \State $C_{ego}\gets C_i \in C \mid a_{ego} \in C_i$ \Comment{Retrieve $a_{ego}$'s community}
            \State $communities \gets communities \cup C_{ego}$
            \State $A \gets V(G) \setminus (C_{ego} \setminus \{a_{ego}\})$ 
            \State $R \gets \{(a_{ego}, a_i) \mid a_i \in A\}$
            \State$G=(A, R, w)$ 
        \Until{ $C_{ego} \setminus a_{ego} = \emptyset $}
        \State \Return $communities$
        \EndProcedure
    \end{algorithmic}
\end{algorithm}

\subsection{Research Group Identification}\label{sec:research-group-detection}
In addition to the community detection step, we introduce a procedure to identify the members of the ego node's research groups over time. The rationale behind this approach is that research groups should not be interpreted solely as sets of relevant co-authors, but rather as temporally coherent subsets of collaborators whose contributions to the ego node's scientific production are relevant and structurally embedded within the collaboration network.

The procedure operates at the level of individual collaborators and is applied annually. For each year, we evaluate the impact of each co-author on the scientific production of $a_{ego}$ by comparing the number of research products co-authored with that collaborator in the given year, against the total number of research products produced by $a_{ego}$ in the same year. This yields a yearly relevance score for each co-author. If such a score exceeds a predefined threshold, then the co-author is considered a member of the ego node's research group in that specific year.

Formally, let $P_{ego}^{\gamma}$ be the set of research products authored by $a_{ego}$ in year $\gamma$. We define $G^\gamma_{ego}$ as:

\begin{equation*}
    G^\gamma_{ego} = (A^\gamma_{ego}, R^\gamma_{ego}, w^\gamma_{ego})
\end{equation*}
where 
\begin{equation*}
    A^\gamma_{ego} = \{a_i \in A_{ego} \mid (a_{ego}, a_i) \in R^\gamma_{ego}\}
\end{equation*}
and
\begin{equation*}
    R^\gamma_{ego} = \{(a_i, a_{ego}) \mid a_{ego} \in auth(p) \wedge a_i \in auth(p) \wedge a_{ego} \neq a_i \wedge p \in P^\gamma_{ego} \}
\end{equation*}
consequently
\begin{equation*}
    w^\gamma_{ego}(a_{ego}, a_i) = \bigl|\{p\in P^\gamma_{ego} \mid a_{ego} \in auth(p) \wedge a_i \in auth(p) 
    \}\bigr| 
\end{equation*}

Let $P_{ego,a_i}^{\gamma} \subseteq P_{ego}^{\gamma}$ be the subset of those products coauthored with author $a_i$.
The yearly impact of author $a_i$ on $a_{ego}$ in year $\gamma$ is defined as:

\begin{equation*}
I(a_i,\gamma)=\frac{|P_{ego,a_i}^{\gamma}|}{|P_{ego}^{\gamma}|}
\end{equation*}

An author $a_i$ is considered a research-group member in year $\gamma$ if:

\begin{equation*}
I       (a_i,\gamma) \geq \tau
\end{equation*}

where $\tau$ is a predefined threshold (30\%). In this way, the method identifies, year by year, the set of collaborators whose contribution represents a substantial fraction of the ego node's yearly scientific output.

Once the yearly members have been identified, the previously extracted community structure from the ego network is used to organise them into distinct research groups. In particular, a research group is assumed to lie entirely within a single community, while its composition may evolve over time. Therefore, for each year, the identified members are grouped according to the community to which they belong. This step allows us to move from a flat list of relevant collaborators to a structured representation in which members are partitioned into distinct community-based groups, which can be further enhanced based on the topics studied in each group or on their affiliations.

Starting with this yearly community-based assignment, research groups are assembled over time by linking groups that exhibit temporal continuity. More specifically, for each year and each community, the set of yearly members is considered a candidate group, provided its size is at least the minimum admissible cardinality. If no valid community-specific grouping can be obtained for a given year, a fallback group containing all yearly members is considered whenever its size is sufficient.

The temporal assembly step proceeds chronologically. For each community, a group detected in year $\gamma$ is connected to the corresponding group observed in a later year if continuity holds. Continuity is verified either when the temporal gap between the two observations does not exceed a fixed tolerance or when the two yearly instances remain structurally connected according to the underlying temporal collaboration information. Whenever continuity is satisfied, the two instances are merged into a single research group, whose temporal span is extended and whose membership is updated accordingly. Otherwise, the previous group is closed, and a new one is started.

At the end of the process, all temporally assembled groups are filtered by minimum size and deduplicated: if two groups share exactly the same set of members, they are merged by extending their temporal intervals to cover the entire observed period. The final output is therefore a chronologically ordered collection of research groups, each represented by its time span and by the set of authors that participated in it.

This procedure enables the identification of research groups as temporally persistent, community-coherent research-group structures around the ego node. Unlike a purely score-based ranking of collaborators, the proposed method captures each co-author's annual relevance and provides a more faithful representation of the ego node's collaborative environment.

\section{The Researcher Group Viewer}\label{sec:tool}
The Research Group Viewer\footnote{The web application is currently available at \url{https://w3id.org/rgviewer}} is a web-based tool that allows users to interactively explore and analyse scientific collaboration networks. Using an ORCID identifier, the tool reconstructs a co-authorship network centred on the specified author, retrieving research outputs linked to that author and their co-authors. Once the network is created, it identifies collaboration patterns, highlights research groups, and visualises how these groups change over time. 

There are two main ways to build the network using the system. The first is a rapid mode that operates directly through OpenAIRE’s graph API. When a user inputs an ORCID identifier, the tool queries the OpenAIRE APIs to fetch associated research products, identifies co-authors with verified ORCID identifiers (which may lead to potential gaps in the network since it excludes research products not yet confirmed in their ORCID profiles), retrieves related works, constructs the co-authorship network, and applies community-detection techniques to find research groups. 
This network is then visualised instantly in the tool's interface, making it ideal for exploratory analyses, as users can create a network dynamically without preparing input files in advance. For those who need a more detailed network that includes works whose ORCID is inferred from metadata sources, users can reach out to one of the authors of this article with the corresponding ORCID identifier to request a pre-computed network that also includes the research products with the inferred ORCID. This two-tiered approach balances accessibility and thoroughness, providing casual users with a quick result while offering a more extensive dataset for those with advanced analytical needs. 

The tool also provides a comprehensive visualisation of identified research groups within the co-authorship network. Authors are depicted as nodes, while co-authorship relationships are shown as weighted edges. 
The system organises authors by their community (as described in Section \ref{sec:community-detection}), allowing users to easily spot clusters of collaborators around the main researcher. The main author can be highlighted and placed at the centre of the visualisation, with co-authors arranged based on their community membership. This setup enables users to view the broader collaboration landscape alongside specific research groups related to the targeted scholar. 

Users can explore the network year by year, providing insights into how collaborations evolve, strengthen, weaken, or fade away. For each year, the interface presents relevant co-authorship relationships, along with the number and type of associated products and the authors involved in various research groups or weaker collaborative frameworks. A timeline view summarises the lifespan of each detected research group (Section \ref{sec:research-group-detection}), showing its start and end years and facilitating understanding of the author's collaborative journey over time. This temporal aspect is particularly valuable for investigating the formation, ongoing presence, and evolution of scientific communities. 

Another important feature of the tool is its ability to recognise different types of research outputs. Collaborations are not treated equally; each edge in the network links two authors (and their products) who worked together and is categorised by type, such as publications, datasets, software, or other outputs. This allows the visualisation to reflect not just the intensity of collaboration through edge weight, but also its qualitative aspects. Thus, the graphical interface can display different product types for each edge, enabling users to differentiate collaborations primarily by whether they are focused on traditional publications or on software- or data-centric research outputs. 

Finally, the interface is connected to OpenAIRE Explore. Users can click on network edges to see the research products involved in a specific collaboration and access detailed records in OpenAIRE Explore. Edge interactions can also generate OpenAIRE queries based on the linked authors, while product links take users directly to individual research outcomes. This integration bridges visual analysis with the bibliographic and research-product data, allowing users to transition seamlessly from the abstract network representation to the concrete outputs that underpin the collaboration.

\section{Evaluation and Discussion}\label{sec:eval}
To the best of our knowledge, no gold standard exists for evaluating the output of ego-network-based community detection algorithms. Hence, we conducted a survey to assess the quality of the research communities identified by our approach and how they evolve over time.
We asked 10 researchers to evaluate our approach. For each evaluator, the collaboration graph was constructed using their ORCID to retrieve their production. 
Only research products in which the evaluator appeared as an author, thanks to a matching ORCID identifier, were included. Co-authors are included only if their ORCID identifier was available in the Graph as well. 

The evaluation was designed to capture both the perceived accuracy of automatically generated research groups and the corrections that domain experts considered necessary. Specifically, we asked each participant to inspect their research groups and to provide structured feedback organised into different sections. In the first section, we collect basic information to know their experience (in terms of years of research) and whether their ORCID profile was curated. The second section was the core of the evaluation, focusing on four possible types of correction to the algorithm: addition (whether some authors were missing from the detected research groups, even if the added authors did not have an ORCID at the moment of the publication), removal (whether our algorithm recognized wrong authors belonging to research groups), split (whether a recognized research group was composed of two or more distinct subgroups), and move (whether a group of authors were recognized in the wrong research group, or they should belong also to another research group). In addition to these qualitative corrections, participants provided an overall accuracy score for the generated research groups on a scale from 1 to 10, where 1 indicated very weak and 10 indicated perfect. Finally, we asked each evaluator for additional remarks and their opinion on what characterises research groups, in order to provide useful guidance for future improvements.

We will now analyse and summarise the responses. The evaluators have an average of 19 years of experience, with 2 having more than 25 years of research experience, 4 having more than 20 years and less than 25, 3 having more than 15 years, and 1 having less than 10 years. Among them, 3 have a curated ORCID profile, meaning that they manually add and update their research products into ORCID, while 6 evaluators only accept the automatic suggestions from ORCID. One evaluator does not update their ORCID profile at all. Overall, our approach achieved an average accuracy of approximately 7.8 out of 10 and the median score of 8/10, with 1 evaluator assigning 5, 1 evaluator assigning 7, 7 evaluators assigning 8, and 1 evaluator assigning 9. These results suggest that the proposed method produced meaningful and accurate research groups while still leaving room for improvement.

The qualitative feedback reveals some interesting patterns. First, every participant identified missing researchers in their research groups. However, we decided to include as missing participants also those without an ORCID profile (e.g., older researchers who ended their careers before ORCID launch), which means that, by design, our approach couldn't identify them. Out of the 39 missing researchers, 16 do not appear because they don't have an ORCID identifier, 
while 6 were already recognised in other research groups. 
This information is quite relevant to our approach and to future work investigating approaches that either consider an author even if their ORCID identifier is missing (which can lead to homonymy problems) or leverage other semantic information from the scholarly knowledge graph to disambiguate those authors.

Second, removals were often associated with sporadic collaborations, annual reports, large shared datasets, or publications involving many co-authors. These cases suggest that co-authorship alone may overestimate the strength of a collaboration when the publication context does not correspond to a stable research relationship. 
In the Graph, we can distinguish research product types and detect whether a product is a deliverable, a report, software, or another type. However, since we evaluated researchers in computer-science-related fields, we decided not to filter them out, as they can be relevant and yield key results for other domains (e.g., maps for cartography). Moreover, we would like to conduct a deeper evaluation of researchers across different fields and assess the nature of collaborations and research results within and across domains to determine which works are more relevant for detecting research groups.

Third, two evaluators suggested splitting or merging certain groups, indicating that the algorithm can occasionally fail to distinguish between closely related but distinct research communities, or, conversely, separate groups that experts perceive as not belonging to the same collaboration context. 

Finally, 5 evaluators suggested a move operation. This highlighted cases in which individual researchers were correctly identified but assigned to a group whose thematic or temporal context did not fully align with the expert’s interpretation.

In discussing the additional remarks and potential future work, evaluators underscored the significance of addressing the temporal aspects of collaboration. Some evaluators pointed out that a research group should not be characterised only by the existence of co-authored publications, but also by the duration, continuity and stability of the collaboration. Additionally, a recommended feature from evaluators is the characterisation of research groups through their disciplines of interest or specific research topics as descriptive elements of the group. This inclusion would help differentiate stable, topic-focused research groups from sporadic collaborations. Another suggestion consists of identifying a stable core of the group, the group leaders, the bridging authors across groups, and the institutional or formal nature of the collaboration, for example, when a group emerges around a research grant, a visiting period, or a structured partnership.
Other valuable suggestions included visualising overlapping memberships across different years, enabling direct feedback within the interface, identifying key individuals who connect various groups, and allowing users to modify the criteria used to define group boundaries.

Ultimately, the evaluation demonstrates that the algorithm can assess research groups even when their ORCID profiles haven't been manually curated. However, feedback indicates that detecting research groups should not exclusively depend on structural co-authorship data. Important factors such as sustained collaboration over time, thematic coherence, publication context, and the involvement of bridging researchers should be incorporated into future improvements to the methodology.



\section{Conclusion and future works}\label{sec:conclusion}
The method proposed in this paper constitutes the first step of a broader research agenda aimed at constructing a multidimensional, evidence-based research fingerprint for individual researchers. 
The research group's identification method presented in this study should not be viewed as a direct measure of a researcher's impact or merit, or as a substitute for qualitative peer review. 
Their purpose is to provide the contextual structure and timeline necessary to interpret other aspects of the research fingerprint. 
The same research output or collaboration behaviour can hold different significance depending on whether it originates within a well-established group, is part of a transition to a new research field or institution, or occurs within the framework of a newly formed interdisciplinary team. 
By reconstructing these contexts, we can enhance the evidence base for responsible research assessments while allowing their interpretation to remain within qualitative, context-sensitive evaluation processes.

It is our intention to assess the scalability of the proposed approach to larger researcher populations and its generalisability across scientific disciplines, where collaboration norms, group sizes, and output types vary considerably.
To this end, we plan to run a larger user evaluation involving researchers from a broader career stage heterogeneity and cross-disciplinary backgrounds.

The main limitation of our approach lies in its dependence on the availability and curation of ORCID identifiers. When an ORCID identifier is absent or has not been manually curated, the reconstructed collaboration network may be partial, as co-authors without a verified ORCID identifier are excluded by design. This limitation points to the need for complementary disambiguation strategies that can leverage other metadata signals available in SKGs.

A second limitation concerns the exclusive reliance on co-authorship structure as the basis for group detection. 
As evaluators noted, co-authorship does not always reflect stable or meaningful research relationships: largely collaborative outputs such as shared datasets, deliverables, or massively co-authored publications can introduce spurious links that inflate the perceived strength of a collaboration. 
Future work should investigate the effect of filtering or down-weighting specific research product types depending on the disciplinary context, while avoiding overly restrictive criteria that may discard relevant non-publication research products in fields such as law or the humanities.

Furthermore, each detected group can be enriched with thematic descriptors -- derived from the topics, keywords, and Fields of Science associated with the group's research results -- so as to distinguish stable, topic-focused communities from heterogeneous or occasional collaborations. The identification of a stable core, bridging authors connecting multiple groups, and the institutional or formal context of the collaboration (e.g., a shared research grant or a structured partnership) will be investigated as additional descriptors.

Finally, the current evaluation is limited to researchers in computer-science-related fields, which constrains the generalisability of the results. Collaboration norms, group sizes, authorship conventions, and output types vary considerably across disciplines, and the algorithm's behaviour in, for example, life sciences or social sciences remains to be assessed.

The identification of stable and evolving research groups around the ego node opens several directions for subsequent investigation, briefly outlined hereafter.

Once the collaborative context of a researcher is established, the natural next step is to characterise the role the ego node plays within each identified group. 
This includes distinguishing between leadership roles -- where the researcher drives the group's agenda -- and contributory or peripheral ones, drawing on authorship position, temporal consistency of collaboration, and network centrality within the subgroup.
Similarly, the temporal evolution of research groups provides a natural lens for studying career trajectories. 
Future work will investigate how a researcher's position and influence within their groups shift over time, using group longevity, membership stability, and authorship patterns as proxies for seniority and scientific maturity.

Building on the group structure, we will analyse the heterogeneity of research outputs produced within each group, distinguishing between publications, datasets, software, and other research products, as well as the correlation between the blend of research outputs and the role and seniority of a researcher within the research group.
This will allow a more granular understanding of how different groups specialise in different output types, and how the ego node's contribution profile varies across contexts and career stages.

With groups and roles established, it becomes possible to study how publication behaviour, such as venue selection, open access practices, and data sharing, varies across a researcher's different collaborative contexts, and how these patterns evolve as groups form, consolidate, or dissolve.

The ultimate goal of this research programme is to integrate all of the above dimensions into a unified, queryable researcher fingerprint, grounded in SKGs data and aligned with the responsible research assessment principles advocated by CoARA, DORA, and the Leiden Manifesto. 
Such a fingerprint could serve as a transparent, reproducible complement to qualitative peer review in a variety of evaluation contexts, from institutional assessments to funding allocation and career progression decisions.

\section*{Data Availability}
The code is available on GitHub at 
\url{https://github.com/simone-angioni/research-group-detection.git}.

\section*{Declaration on Generative AI}
During the preparation of this work, the authors used ChatGPT in order to improve the language and readability. After using this tool, the authors reviewed and edited the content as needed and take full responsibility for the content of the published article.
\bibliography{bibliography}

\end{document}